\documentclass[showpacs,twocolumn,tightenlines,prb]{revtex4-1}
\usepackage{graphicx}
\usepackage{amsmath,amsfonts}
\numberwithin{equation}{section}

\graphicspath{{figures/}} 

\newcommand{\bwt}{\begin{widetext}}
\newcommand{\ewt}{\end{widetext}}

\newcommand{\be}{\begin{equation}}
\newcommand{\ee}{\end{equation}}
\newcommand{\bea}{\begin{eqnarray}}
\newcommand{\eea}{\end{eqnarray}}  

\usepackage{hyperref}
\hypersetup{
    colorlinks=true,
    linkcolor=blue,
    filecolor=magenta,      
    urlcolor=blue,
    citecolor=magenta,
}

\usepackage{color} 
\newcommand{\itt}{\it}

\usepackage{comment}
\def\comment#1{}

\begin{document}

\title{Eliashberg theory in the weak-coupling limit: results on the real frequency axis}

\author{Sepideh Mirabi, Rufus Boyack, and F. Marsiglio}
\affiliation{Department of Physics, University of Alberta, Edmonton, Alberta T6G~2E1, Canada}

\begin{abstract}
We formulate and solve the Eliashberg equations on the imaginary frequency axis at temperatures below $T_c$ in the weak-coupling limit. 
We find an excellent scaling at all temperatures, for a given coupling strength, and the normalized order parameter exhibits a BCS-like temperature dependence. 
The hybrid real-imaginary axis equations are also solved to obtain numerically exact analytic continuations from the imaginary frequency axis to the real frequency axis. 
This provides a determination of the gap edge, which, in the weak-coupling limit, is identical to the order parameter from the imaginary axis. 
The analytical result for the zero-temperature gap edge deviates from the BCS result by a factor of $1/\sqrt{e}$, which was also obtained for the transition temperature $T_c$.
We show that the normalized gap function on both the real and imaginary frequency axes, for an electron-phonon Einstein spectrum ($\delta$-function) of a given strength, is a universal function of frequency, independent of temperature.
The $1/\sqrt{e}$ correction is a result of this non-trivial frequency dependence in the gap function.  
This modification, in the gap edge and in $T_{c}$, serves to preserve various dimensionless ratios to their BCS values.
\end{abstract}

\date{\today}
\maketitle

\section{Introduction}

The Eliashberg theory of superconductivity~\cite{eliashberg60} is often regarded as the ``strong-coupling'' version of the Bardeen-Cooper-Schrieffer (BCS) theory of superconductivity.~\cite{bardeen57} 
This label is a misnomer due to the fact that Eliashberg theory is also a weak-coupling theory, namely, it is based upon the existence of an underlying Fermi sea, and moreover the electron-phonon interaction is assumed to be subject to the Migdal approximation.~\cite{migdal58} 
In both theories, however, a range of interaction strengths are possible.  
Historically it has been the case that Eliashberg theory is invoked~\cite{scalapino69,mcmillan69,carbotte90} to explain somewhat anomalous behaviour of certain 
experimental properties in superconducting materials (Pb, Hg) whose interaction strength is stronger than the prototypical BCS superconductor (Al). 
But even more to the point, BCS theory is not even the weak-coupling limit of Eliashberg theory.~\cite{karakozov76,wang13,marsiglio18}
In fact, while various dimensionless ``BCS ratios" calculated in Eliashberg theory do reduce to their BCS limits as the coupling strength is decreased, 
this is {\it not} true for other important properties, such as the transition temperature $T_c$.

In this paper we further explore the weak-coupling limit of Eliashberg theory, extending the analysis to properties other than $T_c$. 
For example, it is well known that the gap ratio, $2\Delta_0/(k_BT_c)$, where $\Delta_0$ is the zero-temperature gap edge in the
single-particle spectrum and will be defined more precisely below, becomes larger in Eliashberg theory than the BCS limit $\sim3.53$, {\it and,} its
value smoothly decreases to this limiting value as the coupling strength is decreased.~\cite{mitrovic84} 
This implies that the zero-temperature gap, $\Delta_0$, must also differ by the same factor of $1/\sqrt{e}$ as has been established for $T_c$.~\cite{karakozov76,wang13,marsiglio18} 
Showing that this is true is one of the purposes of this paper, and to do so will require analytic continuation of the Eliashberg equations to the real 
frequency axis.~\cite{marsiglio88}

Much of the framework for the present study was laid out in Ref.~[\onlinecite{marsiglio18}]. In Sec.~\ref{ImagAxis} we will briefly review the necessary equations required below
the superconducting transition temperature. More extensive discussion can be found in recent reviews.~\cite{marsiglio08,marsiglio19} 
First we will present the gap equations below $T_c$ on the imaginary axis and discuss their weak-coupling solutions. 
As in  Ref.~[\onlinecite{marsiglio18}], we will adopt the simplest model for the electron-phonon interaction, which assumes a wavevector-independent coupling to an Einstein phonon mode. 
It turns out that this means, particularly in the weak-coupling limit, that the gap edge is already determined by these imaginary-axis solutions. 
Nevertheless, it is desirable to compute the full frequency-dependent gap function, and for this we require the real frequency axis equations, which are investigated in Sec.~\ref{RealAxis}. 
We also derive an analytical approximation which produces the remarkable $1/\sqrt{e}$ factor to achieve the desired cancellation in the gap ratio.
In Sec.~\ref{Conc} we conclude with a summary and a discussion of other dimensionless BCS ratios.

\section{Eliashberg Theory on the Imaginary Axis below $T_{c}$}
\label{ImagAxis}

Following the assumptions and approximations discussed in Ref.~[\onlinecite{marsiglio18}], 
the Eliashberg equations are given by~\cite{marsiglio19}

\bwt
\begin{align}
Z_N(i\omega_m) &= 1 + {\pi T \over \omega_m} \left(\lambda + 2 \sum_{n=1}^{m-1} \lambda(i\nu_n) \right).
\label{zz_norm} \\
Z(i\omega_m)  &=  Z_N(i\omega_m) +  {\pi T \over \omega_m} \sum_{m^\prime} \lambda(i\omega_m - i\omega_{m^\prime})
\left( {\omega_{m^\prime} \over  \sqrt{\omega^2_{m^\prime} + \Delta^2(i\omega_{m^\prime})} } - {\rm sgn}(\omega_{m^\prime})\right).
\label{zz_explicit} \\
Z(i\omega_m) \Delta(i\omega_m)  &=  \pi T \sum_{m^\prime}\lambda(i\omega_m - i\omega_{m^\prime})
{\Delta(i\omega_{m^\prime}) \over \sqrt{\omega^2_{m^\prime} + \Delta^2(i\omega_{m^\prime})}}.
\label{gap}
\end{align}

\ewt

Here, $T$ is the temperature, $\omega_m \equiv (2m-1)\pi T$ and $\nu_n \equiv 2n \pi T$ are the Fermion and Boson Matsubara frequencies, respectively, with $m$ and $n$ integers.
Natural units $\hbar=k_{B}=1$ are used throughout the paper; Boltzmann's constant is restored in discussions of the gap ratio. 
The function $Z(i\omega_m)$ is related to the odd part of the electron self energy through~\cite{marsiglio08}
\begin{equation}
i\omega_m \left[ 1 - Z(i\omega_m) \right]   \equiv
{1 \over 2} \left[ \Sigma(i\omega_m) - \Sigma(-i\omega_m) \right]
\label{zz}
\end{equation}
and the even part is identically zero, due to electron-hole symmetry. The so-called gap function is defined as $\Delta(i\omega_m) \equiv  \phi(i\omega_m)/Z(i\omega_m)$, where $\phi(i\omega_m)$ is the so-called pairing function.  
As previously discussed,~\cite{marsiglio18,marsiglio19} these functions are often assumed to have no important wavevector dependence, and here this has been dropped. 
The ``glue''  in these equations is represented by the electron-phonon propagator, contained in
\be
\lambda(i\nu_n)  = {\lambda \omega_E^2 \over \omega_E^2 - (i\nu_n)^2}
\label{lambda_n}
\ee
for the Einstein model mentioned earlier. Here, the constant $\lambda$ is the dimensionless electron-phonon coupling constant and $\omega_E$ is the Einstein phonon frequency. 
As in Ref.~[\onlinecite{marsiglio18}], we omit the direct Coulomb repulsion term.

\begin{figure*}[htp]
\centering
\begin{tabular}{@{}cc@{}cc@{}}
\includegraphics[height=2.3in,width=2.2in,angle=-90]{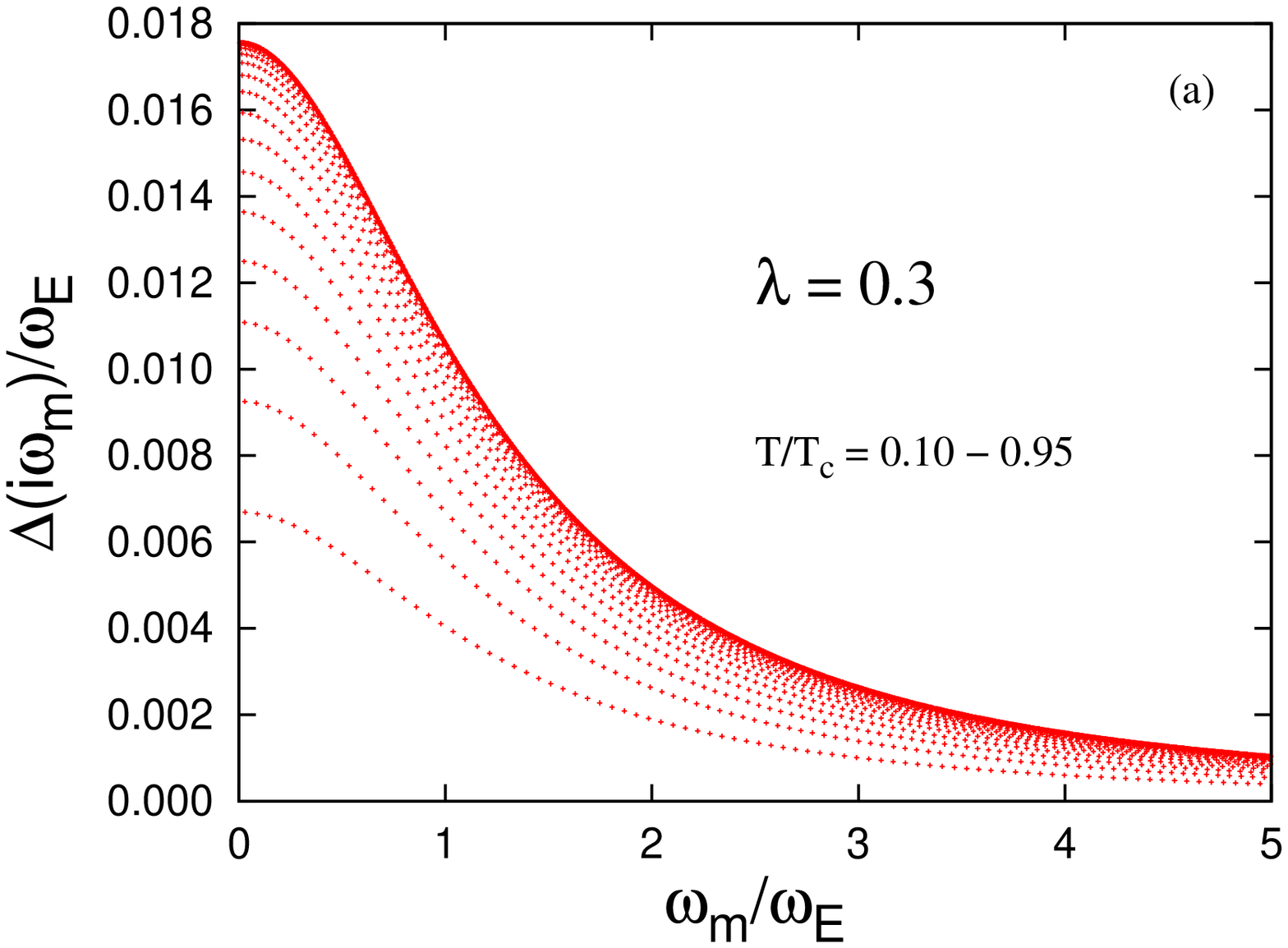} &
\includegraphics[height=2.3in,width=2.2in,angle=-90]{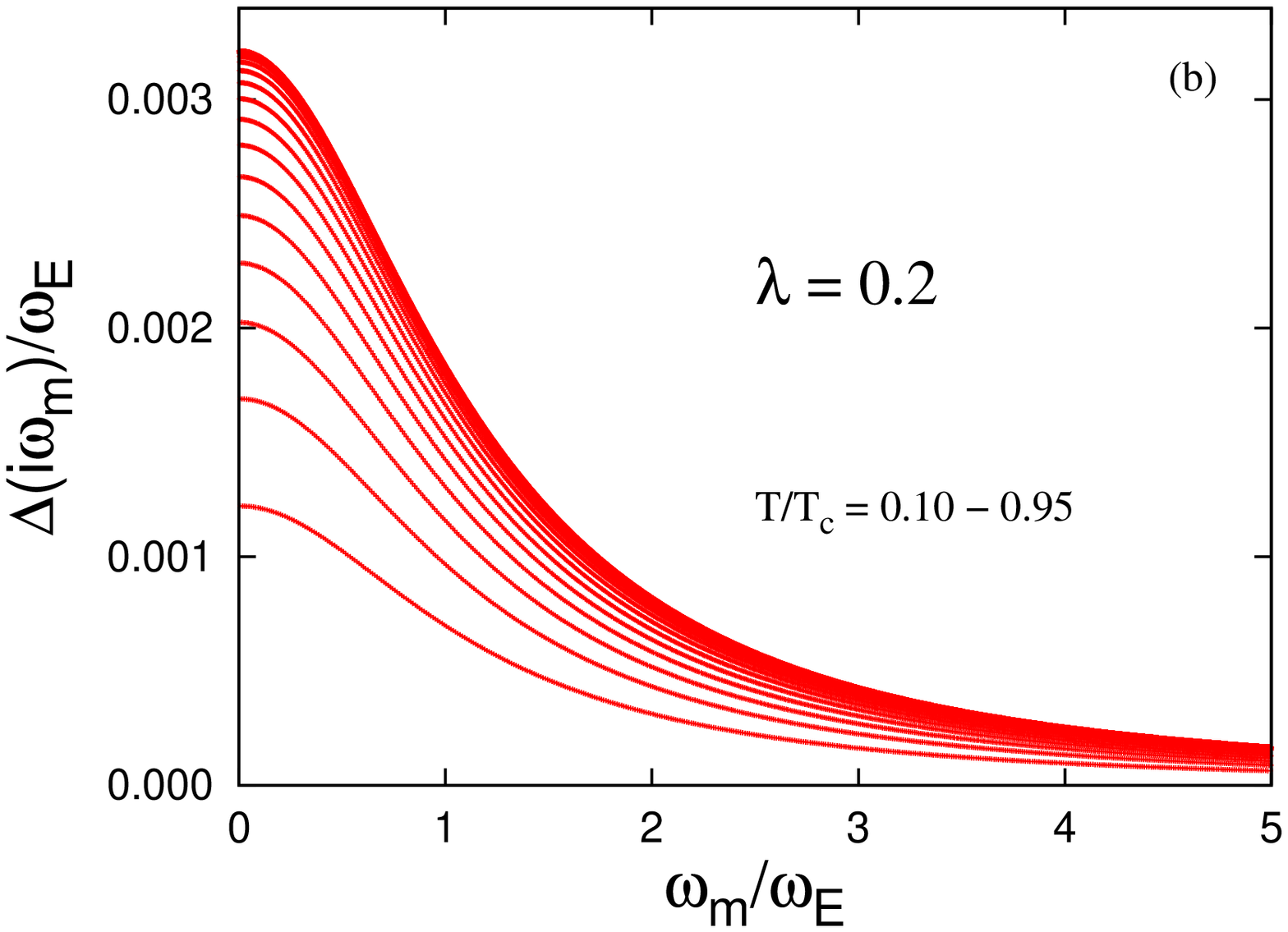} &
\includegraphics[height=2.3in,width=2.2in,angle=-90]{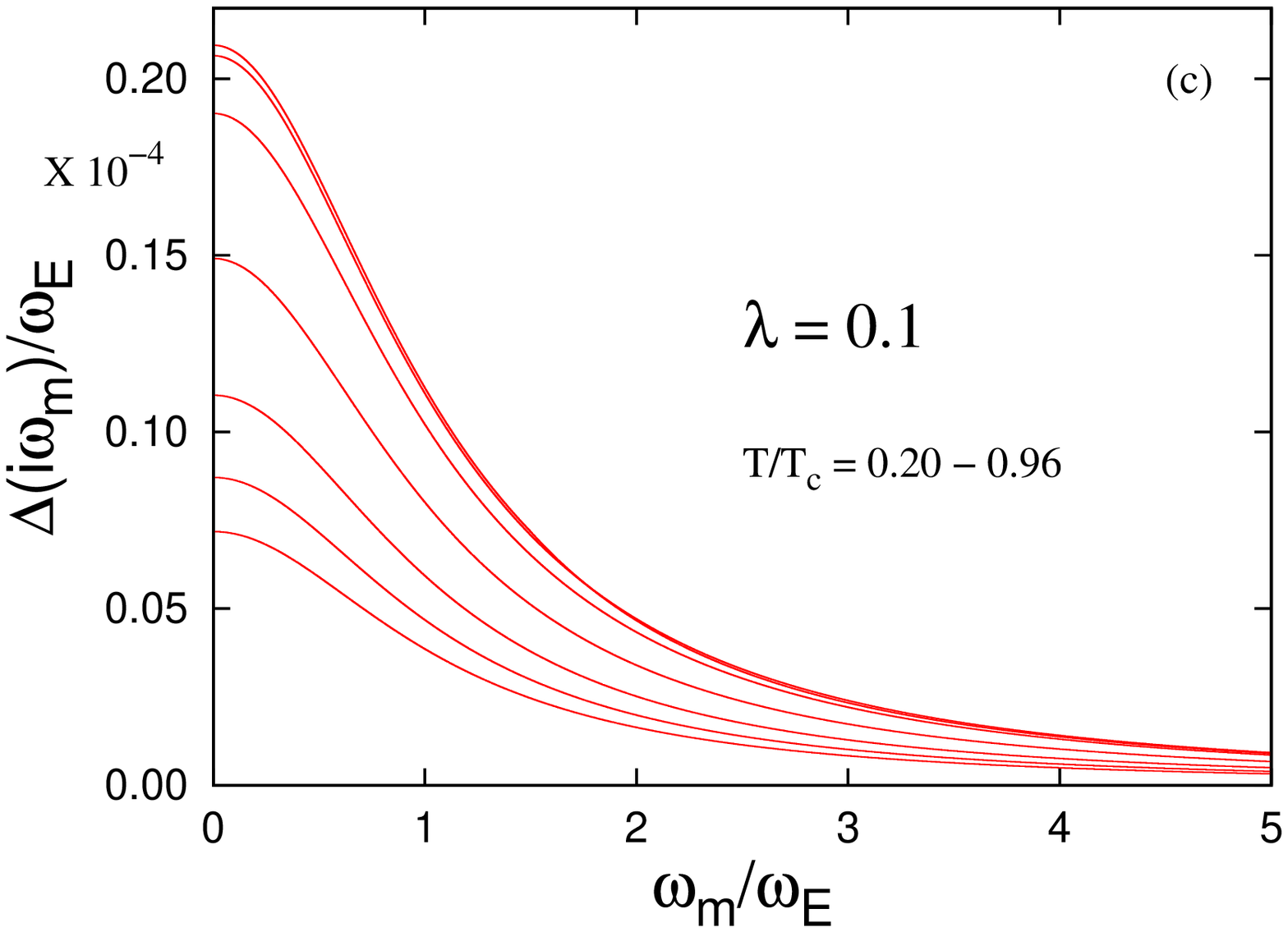}
\end{tabular}
\caption{The gap function on the imaginary axis, $\Delta(i\omega_m)$, as a function of $\omega_m/\omega_E$ for various temperatures below
$T_c$, for (a) $\lambda = 0.3$, (b) $\lambda = 0.2$, and (c) $\lambda = 0.1$. All of these curves (really discrete sets of points) show similar behaviour --
a gradual decrease as the frequency increases. As a function of temperature, in each case the scale increases as the temperature is lowered towards $T=0$. 
As a function of coupling strength, the main difference is simply the scale on the vertical axis, which scales with $T_c$. 
These are  $T_c/\omega_E = 0.009923$ for $\lambda = 0.3$, $T_c/\omega_E = 0.001821$ for $\lambda = 0.2$, and 
$T_c/\omega_E = 0.1189 \cdot 10^{-4}$ for $\lambda = 0.1$. In (a) and (b) we used $T/T_c = 0.10 - 0.95$ in steps of $0.05$ (18 temperatures each),
with the highest curves corresponding to the lowest temperatures. It is clear that many of the low-temperature results in each case are almost
identical. In (c) we used $T/T_c = 0.20, 0.60, 0.80, 0.90, 0.94, 0.96$.}
\label{fig1_mirabi}
\end{figure*}

The numerical and analytical solutions to the linearized gap equations have already been discussed in Ref.~[\onlinecite{marsiglio18}]. 
As is clear from Fig.~(3b) in Ref.~[\onlinecite{marsiglio19}], the frequency dependence of the gap function, on the imaginary axis,  is not so different at temperatures below $T_c$ from the frequency dependence at $T_c$. 
To numerically solve the Eliashberg equations, we have adopted a simple iteration scheme that works very efficiently at low temperatures, 
particularly when the iteration is initiated using the frequency dependence at $T_c$. 
Convergence for temperatures close to $T_c$ is problematic, and our simple-minded approach sometimes requires more than 1,000 iterations to converge, as opposed to 10-20 iterations at low temperatures. 
Nonetheless, everything progresses fairly quickly (on a laptop) and we did not exert extra effort to improve this ``slowdown.'' 

We first show numerical results for the gap function, in units of $\omega_E$, on the imaginary axis as a function of Matsubara frequency. This is plotted in
Fig.~(\ref{fig1_mirabi}) for a variety of temperatures, with progressively weaker couplings, (a) $\lambda = 0.3$, (b) $\lambda = 0.2$, and (c) $\lambda = 0.1$. 
These figures look very similar to one another; the main difference is the scale of the gap function, since $T_c$ changes considerably as a function of $\lambda$. 
Note that the low-frequency gap function [$\Delta(i\omega_{m=1})$ for definiteness] acts as an order parameter for the superconducting phase transition --  it is zero at $T_c$ and increases to its full value at $T=0$. 
The temperature dependence is BCS-like, and was shown in Ref.~[\onlinecite{marsiglio19}] to be indistinguishable from BCS already for $\lambda = 0.3$.

\begin{figure}[tp]
\begin{center}
\includegraphics[height=3.2in,width=2.8in,angle=-90]{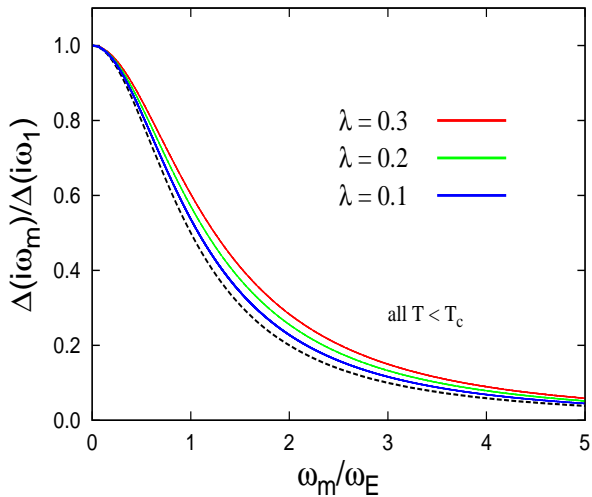}
\end{center}
\caption{The normalized gap function,  $\Delta(i\omega_m)/\Delta(i\omega_1)$, versus $\omega_m/\omega_E$ for each temperature shown in Fig.~(\ref{fig1_mirabi}).
Each of the three curves correspond to a particular coupling strength, as indicated. In fact, these plots consist of many different curves corresponding to the temperatures shown in Fig.~(\ref{fig1_mirabi}). 
For a given weak-coupling strength, the frequency dependence is universal. 
Also shown (dashed black curve) is the limiting result for $\lambda \rightarrow 0$, given by $1/[1 + (\omega_m/\omega_E)^2]$.}
\label{fig2_mirabi}
\end{figure}

We now address the imaginary-axis frequency dependence as a function of temperature for a given coupling strength.
In Fig.~(\ref{fig2_mirabi}), we show the normalized gap function $\Delta(i\omega_m)/\Delta(i\omega_1)$ versus $\omega_m/\omega_E$ for each temperature and coupling strength.
This normalized gap function will go to unity at low frequency by definition. 
Nonetheless, we emphasize that all of the curves from Fig.~(\ref{fig1_mirabi}) (more than 40 curves) are shown in Fig.~(\ref{fig2_mirabi}). 
Every normalized gap function for a given coupling strength, irrespective of the temperature, collapses onto the same curve, as indicated by the fact that only one curve is visible for each coupling strength.
The fourth curve shown is the result expected in the weak-coupling limit ($\lambda \rightarrow 0$) and is given by $\Delta_{\rm weak}(i\omega_m)/\Delta(i\omega_1) = 1/[1 + (\omega_m/\omega_E)^2]$. 
It is evident that the results are systematically trending towards this result, as the coupling strength is decreased towards zero.
This means, of course, that the required number of iterations for convergence should be minimal; nonetheless, as mentioned earlier, we found that a very large number of iterations was still required for temperatures close to $T_c$, whereas relatively few iterations 
were needed at lower temperatures.~\cite{remark1}

The absence of any discernible temperature dependence in Fig.~(\ref{fig2_mirabi}) implies a profound result. Indeed, following Ref.~[\onlinecite{marsiglio18}] we derived an 
approximation for the gap function based on a weak-coupling expansion in $\lambda$, for all temperatures below $T_c$, to complement the expression already obtained at $T_c$. 
However, given the result in Fig.~(\ref{fig2_mirabi}), it is clear that we obtain precisely the same expression as that at $T_c$, given by Eq.~(32) and Eq.~(38) in Ref.~[\onlinecite{marsiglio18}]. 
The reason for this is because the $\Delta(i\omega_m)/\omega_E$ terms appearing in the square root in Eqs.~(\ref{zz_explicit}-\ref{gap}) are exponentially small in $1/\lambda$, as illustrated in Fig.~(\ref{fig1_mirabi}) and will be proven rigorously in the next section, 
and similarly the terms of order $T_c/\omega_E$ are also exponentially small in $1/\lambda$. 
Thus, the weak-coupling form of the frequency-dependent gap, on the imaginary axis, has the same functional form, irrespective of the temperature. 
We will elaborate further on this when we examine the real axis expression.

\section{Real frequency axis results}
\label{RealAxis}

Determining the gap ratio, $2\Delta_0/(k_BT_c)$, strictly speaking requires knowledge of the gap function $\Delta(\omega + i\delta)$, which is obtained by analytical continuation of the imaginary frequency axis results to just above the real frequency axis. 
In point of fact, the ``gap'', $2\Delta_0(T)$ (called $2\epsilon_0$ in Ref.~[\onlinecite{bardeen57}]), has been defined historically through the single-particle density of states:
\begin{equation}
{g(\omega) \over g(\epsilon_F)}  = {\rm Re}\left[{\omega \over \sqrt{(\omega + i\delta)^2 - \Delta^2(\omega + i\delta)}}\right],
\label{dos}
\end{equation}
where $g(\epsilon_F)$ is the (assumed) constant density of states at the Fermi level. 
Given that, within the BCS approximation, the gap function is a constant, $\Delta(\omega + i\delta) = \Delta_{\rm BCS}$, and that the square-root in Eq.~(\ref{dos}) should have the same sign as $\omega$, 
then the single-particle density of states shows a clear gap above the Fermi level (at any temperature) given by $\Delta_0 = \Delta_{\rm BCS}$. 
Within Eliashberg theory, however, the argument is more nuanced, as first noticed by Karakozov et al.~\cite{karakozov76} 
The low-frequency dependence of the gap function is very subtle, with different frequency dependence at $T=0$ compared with $T> 0$. 
In the weak-coupling limit, the frequency regime in which ``gapless'' superconductivity is possibly relevant is exponentially small, 
and thus  unobservable, so we proceed with the above definition; for Eliashberg theory the gap is thus determined from the condition
\begin{equation}
\Delta_0 \equiv {\rm Re}[\Delta(\omega = \Delta_0)].
\label{gap_defn}
\end{equation}
 A description of the regime in which ``gapless'' superconductivity is more visible is given in Ref.~[\onlinecite{marsiglio91}].

Determining this gap requires us to calculate the gap function on the real frequency axis.~\cite{remark2} For an Einstein phonon spectrum, in the weak-coupling limit
it will turn out that the imaginary axis result suffices, but we will describe the full procedure to keep this description general. 
For the Einstein phonon spectrum used here, the equations that analytically continue the imaginary frequency axis results to the real frequency axis are~\cite{marsiglio88}

\bwt
\begin{align}
\phi(\omega + i \delta)  &= \pi T\sum_{m=-\infty}^{\infty}
\lambda(\omega-i\omega_m){\Delta({i\omega_m}) \over \sqrt{\omega_m^2 + \Delta^2({i\omega_m})}}\nonumber \\
 &\quad+  i\pi \lambda{ \omega_E \over 2}  \Biggl\{\left[N(\omega_E)+f(\omega_E-\omega)\right] 
 {\phi(\omega-\omega_E + i \delta) \over \sqrt{(\omega - \omega_E + i\delta)^2 Z^2(\omega-\omega_E + i \delta) - \phi^2(\omega-\omega_E + i \delta)}}\nonumber \\
 &\hspace{1.7cm} + \left[N(\omega_E)+f(\omega_E+\omega)\right] {\phi(\omega+\omega_E + i \delta) \over \sqrt{(\omega + \omega_E + i\delta)^2 Z^2(\omega+\omega_E + i \delta) - \phi^2(\omega+\omega_E + i \delta)}}\Biggr\},
 \label{f1}\\
Z(\omega + i\delta)  &=  1 + {i\pi T \over \omega}\sum_{m=-\infty}^{\infty} 
\lambda(\omega-i\omega_m) {\omega_m  \over \sqrt{\omega_m^2  + \Delta^2({i\omega_m})} } \nonumber \\
 &\hspace{0.6cm}+  i\pi \lambda{ \omega_E \over 2\omega} \Biggl\{\left[N(\omega_E)+f(\omega_E-\omega)\right]
  {(\omega-\omega_E)Z(\omega-\omega_E + i \delta) \over \sqrt{(\omega-\omega_E + i\delta)^2 Z^2(\omega-\omega_E + i \delta) - \phi^2(\omega-\omega_E+i\delta)}} \nonumber \\ 
&\hspace{2cm}  + \left[N(\omega_E)+f(\omega_E+\omega)\right]{(\omega+\omega_E)Z(\omega+\omega_E + i \delta)
\over \sqrt{(\omega+\omega_E + i\delta)^2 Z^2(\omega+\omega_E + i \delta) - \phi^2(\omega+\omega_E + i \delta)}}\Biggr\}.
\label{f2}
\end{align}
\ewt

Here, $ \phi(\omega + i\delta) \equiv Z(\omega + i\delta) \Delta(\omega + i\delta)$, and $f(\omega) \equiv 1/({\rm exp}(\omega/T) + 1)$ and
$N(\nu) \equiv 1/({\rm exp}(\nu/T) - 1)$ are the Fermi-Dirac and Bose-Einstein distribution functions respectively. 
In Eqs.~(\ref{f1}-\ref{f2}), the branch of the square root with positive imaginary part is used.~\cite{marsiglio19} 
We iterate these equations only on the positive real frequency axis and use symmetries~\cite{marsiglio19} 
to relate functions with a negative real argument to their positive real-argument counterpart. 
In the weak-coupling limit, if we are interested only in low-frequency properties (such as the gap defined above), then only the first line in each equation needs to be retained, 
as the second and third lines are exponentially small in $1/\lambda$ (recall that the third line in each expression is identically zero at $T=0$ for all coupling strengths).~\cite{marsiglio88}
This is just the result we would obtain if we na\"ively (and generally incorrectly)
replaced the Matsubara frequency $i\omega_m$ in Eqs.~(\ref{zz_norm},\ref{zz_explicit},\ref{gap}) with the real-axis frequency $\omega + i\delta$. 
We proceed with the theoretical analysis in this fashion and sketch some approximations, following those presented at $T_c$ in Ref.~[\onlinecite{marsiglio18}].
As on the imaginary axis, the low-frequency dependence of the renormalization function is $Z(\omega+i\delta) \approx 1 + \lambda$; 
see the discussion in Ref.~[\onlinecite{marsiglio19}] concerning the low-frequency behaviour of ${\rm Im} Z(\omega + i\delta)$, which can be ignored here.

\bwt
For the gap equation, as argued above we focus only on the first line of Eq.~(\ref{f1}) and rewrite the kernel as
\begin{equation}
\lambda(\omega-i\omega_m)=
{\lambda \omega_E^2 \over \omega_E^2 + (\omega_m + i\omega)^2} = {\lambda \over 1 + (\bar{\omega}_m + i\bar{\omega})^2}
= \lambda \left[ {1 \over 1 - \bar{\omega}^2} + \left( {1 \over 1 + (\bar{\omega}_m + i\bar{\omega})^2} - {1 \over 1 - \bar{\omega}^2}  \right) \right]
\label{kernel}
\end{equation}
where we define $\bar{Q} \equiv Q/\omega_E$. Thus, it is apparent from Eq.~(\ref{f1}) and Eq.~(\ref{kernel}) that the approximate gap function, denoted $\Delta_{\rm app}$, can be written to first order in $\lambda$ as
\ewt
\begin{equation}
Z^{(1)}_{\rm app}(\omega + i\delta)\Delta^{(1)}_{\rm app}(\omega + i\delta) = {\Delta_0 \over 1 - \bar{\omega}^2}\left(1 + \lambda h(\omega) \right).
\label{pert}
\end{equation}
Equation~(\ref{pert}) is the real-axis analogue of Eq.~(18) in Ref.~[\onlinecite{marsiglio18}], 
where $h(\omega)$ has replaced $f(\omega_m)$ and the  gap function now has
amplitude $\Delta_0$ since we are considering the problem below $T_c$. We have momentarily dropped the $i\delta$ term -- it is understood to accompany
$\omega$ everywhere (although at times it is unnecessary, for instance when it appears in Fermi functions). 
Note that the $\lambda \rightarrow 0$ limit is clearly recognizable as $1/(1 + \bar{\omega}_m^2) = 1/(1 - (i\bar{\omega}_m)^2)$ is analytically continued to $\rightarrow 1/(1 - (\bar{\omega} + i\delta)^2)$. 
The expression for $h(\omega)$, however, is different from its imaginary axis counterpart (Eq.~(32) in Ref.~[\onlinecite{marsiglio18}]), and is given by
\bwt
\begin{align}
h(\omega) &= {3 \over 2} -{1 \over 4\bar{\omega}}\ {\rm Re}\biggl[ \psi\left({1 \over 2} + i\left({\omega_E + \omega \over 2\pi T}\right)\right)  - 
\psi\left({1 \over 2} + i\left({\omega_E - \omega \over 2\pi T}\right)\right] \nonumber \\
&\quad\quad - {3 \over 4}{1 \over 2 - \bar{\omega}} \ {\rm Re}\left[\psi\left({1 \over 2} + i{\omega_E \over 2\pi T}\right) - 
 \psi\left({1 \over 2} + i\left({\omega_E - \omega \over 2\pi T}\right)\right)\right] \nonumber \\
&\quad\quad - {3 \over 4}{1 \over 2 + \bar{\omega}} \ {\rm Re}\left[\psi\left({1 \over 2} + i{\omega_E \over 2\pi T}\right)  - 
 \psi\left({1 \over 2} + i\left({\omega_E + \omega \over 2\pi T}\right)\right)\right].
\label{hfunc}
\end{align}
Further simplifications are possible for the limits $T\ll\omega_{E}$ or $\omega\ll\omega_E$.  In the weak-coupling limit ($\lambda\ll1$), $T_{c}\ll\omega_{E}$ and therefore it 
follows that $T\ll\omega_{E}$. 
Using the asymptotic expansion of the digamma function,~\cite{ASBook} $\psi\left(z+\frac{1}{2}\right)\approx \ell n(z)+O(z^{-2})$, as $|z|\rightarrow\infty$, the result for $h(\omega)$ is 
\be
h(\omega) \approx {3 \over 2}   - {1 \over 4 - \bar{\omega}^2}\left[ {2 + \bar{\omega}^2 \over 2\bar{\omega}}{\rm \ell n}\left|  {1 + \bar{\omega}   \over 1 - \bar{\omega} } \right|  - {3 \over 2} {\rm \ell n}\left|1 - \bar{\omega}^2\right| \right].
\label{hasym}
\ee
Technically, this expression is valid only in the limits $\omega_{E}/T\gg1$ and $(\omega_{E}-\omega)/T\gg1$; that is, at $\omega=\omega_{E}$ it is invalid. 
The exact result for $h(\omega=\omega_{E})$, in the limit $\omega_{E}\gg T$, is given by 
\be
h(\omega=\omega_{E}) \approx \frac{3}{2}-\log\left({\omega_{E} \over 2\pi T}\right)+\psi\left({1 \over 2}\right),\quad \omega_{E}\gg T.
\ee
The expression in Eq.~(\ref{hfunc}) is very different from the imaginary-axis expression, given in Eq.~(32) combined with Eq.~(38) in Ref.~[\onlinecite{marsiglio18}]:
\be
f(\omega_m) \approx {3 \over 2}   - {1 \over 4 + \bar{\omega}_m^2}\left[
{2 - \bar{\omega}_m^2 \over \bar{\omega}_m}{\rm tan}^{-1}\bar{\omega}_m  - {3 \over 2} {\rm \ell n}(1 + \bar{\omega}_m^2) \right],
\label{ffz}
\ee
but clearly Eq.~(\ref{hasym}) is the natural analytic continuation of Eq.~(\ref{ffz}).
Following the analysis in Ref.~[\onlinecite{marsiglio18}], and using only the first line in Eq.~(\ref{f2}), the weak-coupling form of the renormalization function, on the real frequency axis, is
\be
Z^{(1)}_{\rm app}(\omega+i\delta)=1+\lambda{1 \over 2\bar{\omega}}{\rm Re}\left[\psi\left({1 \over 2} - i\left({\omega_E + \omega \over 2\pi T}\right)\right)-\psi\left({1 \over 2} + i\left({\omega_E - \omega \over 2\pi T}\right)\right)\right].
\ee
\ewt
In the limit $\omega_{E}/T\gg1$ and $(\omega_{E}-\omega)/T\gg1$, this result reduces to 
\begin{equation}
Z^{(1)}_{\rm app}(\omega+i\delta) \approx 1 + \lambda{1  \over 2\bar{\omega}}{\rm \ell n} \left|{1 + \bar{\omega} \over 1 - \bar{\omega}}\right|.
\label{zzapp_real}
\end{equation}
By contrast, the result on the imaginary frequency axis has the limiting form
\begin{equation}
Z(i\omega_m) \approx 1 + \lambda {1 \over \bar{\omega}_m} {\tan}^{-1} \bar{\omega}_m,
\label{zzapp}
\end{equation}
obtained by ignoring terms of order $(T_c/\omega_E)^2$, which are exponentially small in $1/\lambda$ in the weak-coupling limit. 
This is also the result obtained in the zero-temperature limit, so it is not too surprising that the analytical continuation ($i\omega_m \rightarrow \omega + i\delta$) of Eq.~(\ref{zzapp}) is in fact the zero-temperature limit of the real-axis expression,
where again we have dropped the $i \delta$, which would in fact contribute an imaginary part for $|\omega| > \omega_E$. 

\begin{figure}[tp]
\begin{center}
\includegraphics[height=3.2in,width=2.8in,angle=-90]{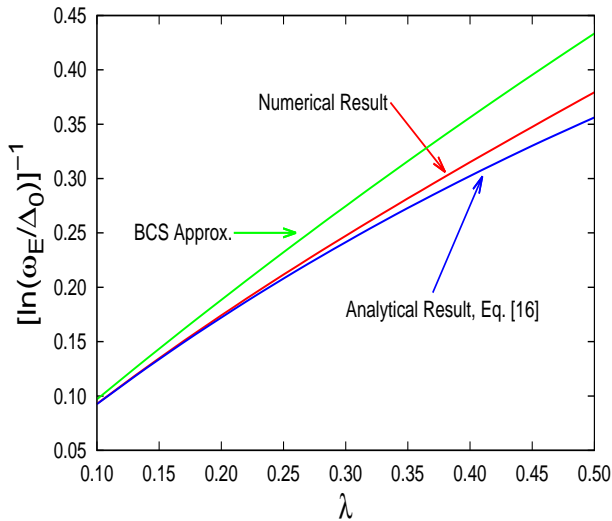}
\end{center}
\caption{A comparison of the various calculations of $[{\rm ln}(\omega_E/\Delta_0)]^{-1}$ versus $\lambda$. 
Numerical results are shown in red. The usual BCS approximation is given by the green curve, while the improved estimate  given by Eq.~(\ref{delta0}) is shown in blue.
This latter result becomes essentially exact for $\lambda {{ \atop <} \atop {\approx \atop }} 0.2$, and the improvement is very similar to
that obtained in Ref.~[\onlinecite{marsiglio18}] for the corresponding quantity with $T_c$ instead of $\Delta_0$.}
\label{fig3_mirabi}
\end{figure}
Analogous to the calculation  of $T_c$, the actual gap at $T=0$ can be derived analytically by substituting Eq.~(\ref{kernel}) into the first line of Eq.~(\ref{f1}). 
A brief derivation is given in the Appendix. The result is
\begin{equation}
\Delta_0 = {2 \omega_E \over \sqrt{e}} {\rm exp}\biggl(-{1 + \lambda \over \lambda} \biggr),
\label{delta0}
\end{equation}
and the same $\sqrt{e}$ appears in the denominator here as in the $T_c$ equation, with the consequence that the weak-coupling limit of the gap ratio is $2\Delta_0/(k_BT_c) = 3.53$, the same as in BCS theory. 
In Fig.~(\ref{fig3_mirabi}) we show the numerical data alongside the BCS and improved approximations. 
The result with the factor of $1/\sqrt{e}$ due to the frequency dependence of the coupling is very accurate in comparison to the simple BCS result. 
Over the entire range of $\lambda$ values shown, $2\Delta_0/(k_BT_c)$ varies between $3.63$ and $3.53$.

\begin{figure*}[tp]
\begin{center}
\includegraphics[height=2.8in,width=2.4in,angle=-90]{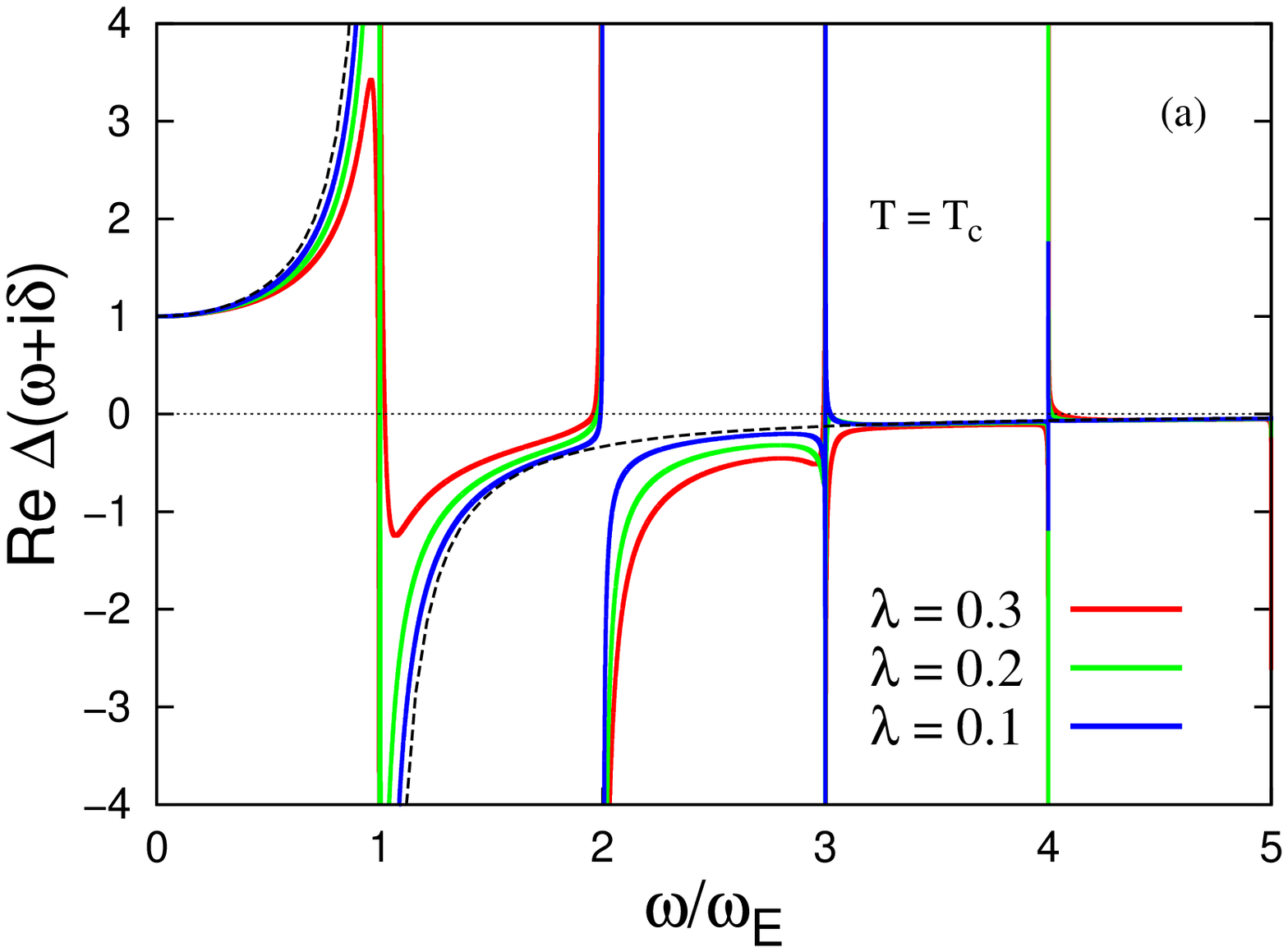}
\includegraphics[height=2.8in,width=2.4in,angle=-90]{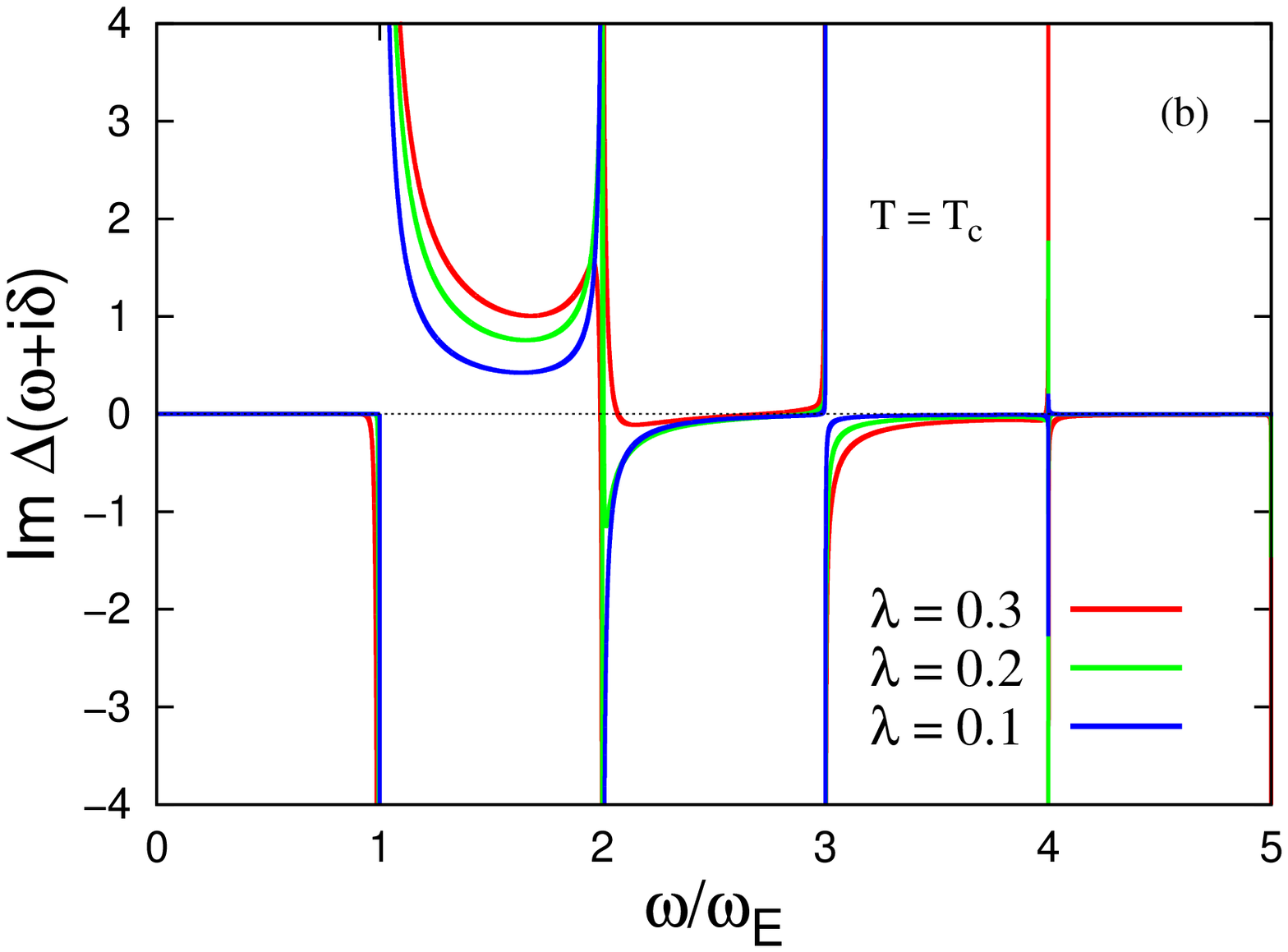}
\end{center}
\caption{(a) Real and (b) Imaginary parts of the gap function, $\Delta(\omega + i\delta)$, at $T_c$, as a function of frequency, for three different coupling strengths, $\lambda = 0.3, 0.2, 0.1$. 
Also shown is the $\lambda \rightarrow 0$ analytical result, $1/[1 - (\omega/\omega_E)^2]$, indicated with a dashed black curve. 
Clearly the numerical results for the real part of the gap function are trending towards this curve. 
Both real and imaginary parts show a series of resonances with decaying amplitudes at integer multiples of the phonon frequency $\omega_E$. 
The existence of a non-zero imaginary part at frequencies $\omega < \omega_E$ is visible in the case of $\lambda =0.3$, and is due to the not-insignificant critical temperature in this case. 
Most notably, as was the case on the imaginary frequency axis, the gap function has a significant frequency dependence, in contrast to the case assumed in standard BCS theory. 
These results are at $T_c$ and hence the gap function is normalized to the low frequency gap function value.~\cite{remark3}}
\label{fig4_mirabi}
\end{figure*}

\begin{figure*}[tp]
\begin{center}
\includegraphics[height=2.8in,width=2.4in,angle=-90]{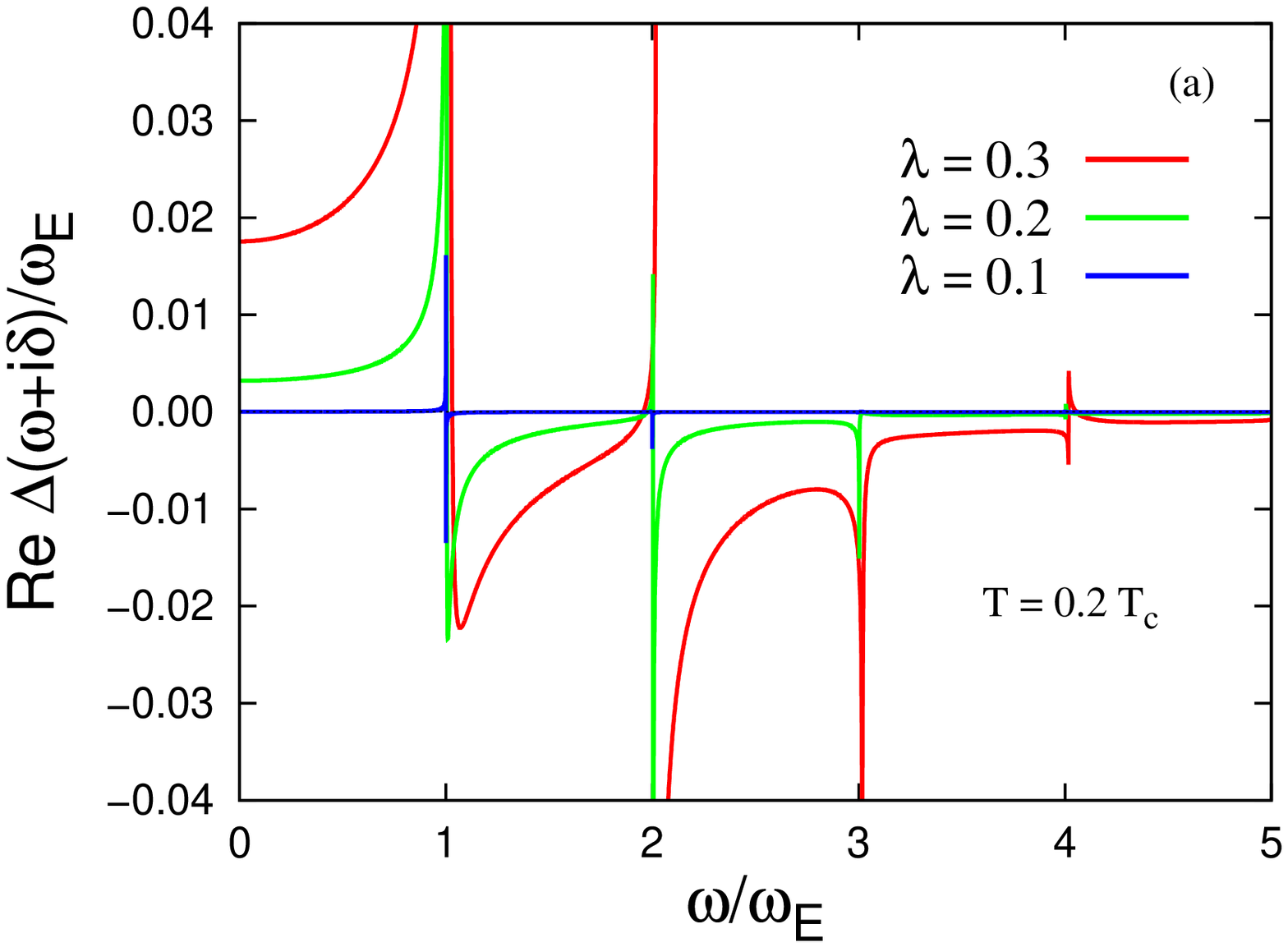}
\includegraphics[height=2.8in,width=2.4in,angle=-90]{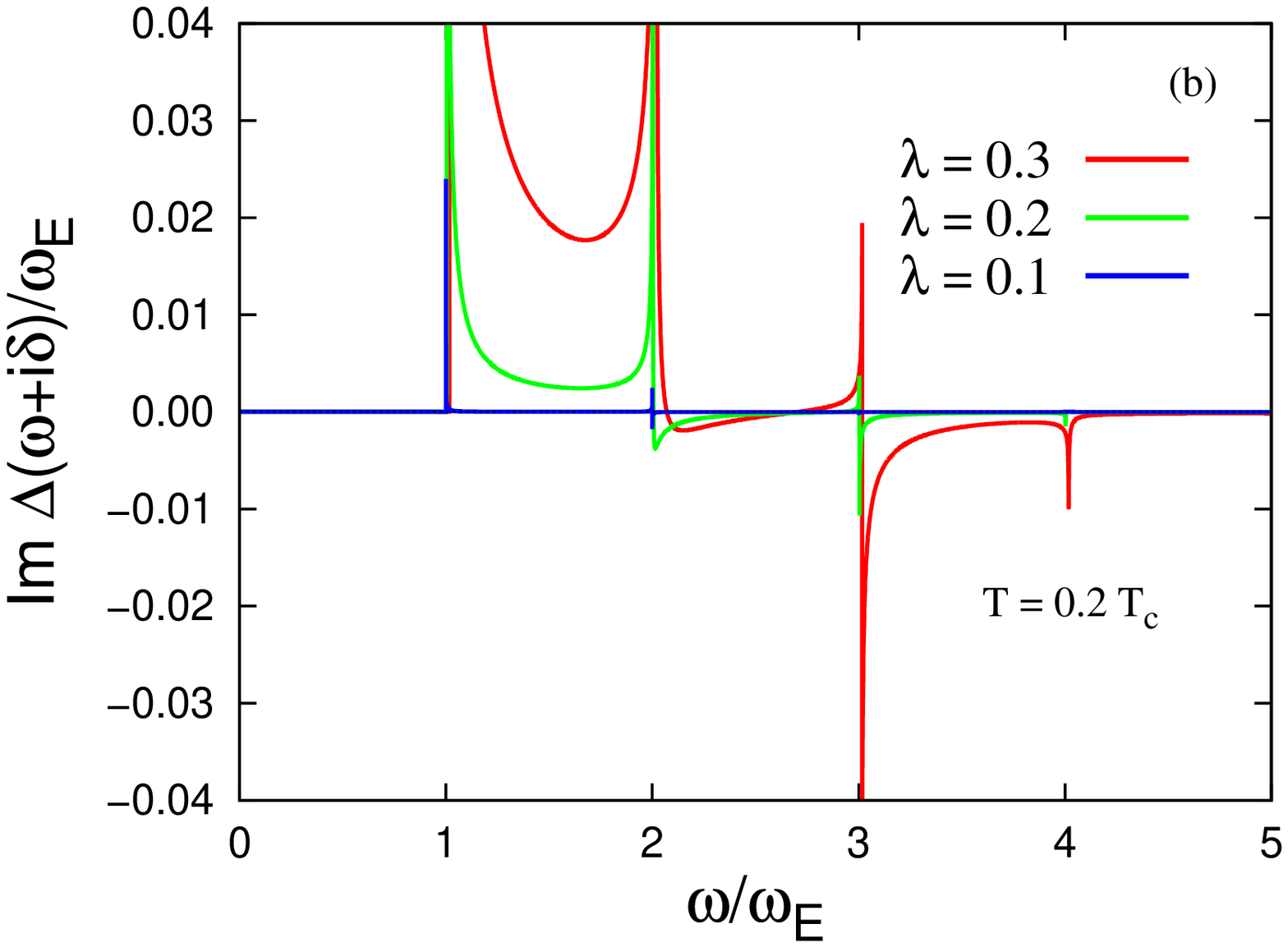}
\end{center}
\caption{(a) Real and (b) Imaginary parts of the gap function, $\Delta(\omega + i\delta)$, at a low temperature $T = 0.2T_c$, as a function of frequency, 
for three different coupling strengths, $\lambda = 0.3, 0.2, 0.1$. Everything is similar to the result in Fig.~(\ref{fig4_mirabi}), but now, because these are
shown on one absolute scale, the results for $\lambda = 0.1$ are barely visible.}
\label{fig5_mirabi}
\end{figure*}

\begin{figure*}[tp]
\begin{center}
\includegraphics[height=2.8in,width=2.4in,angle=-90]{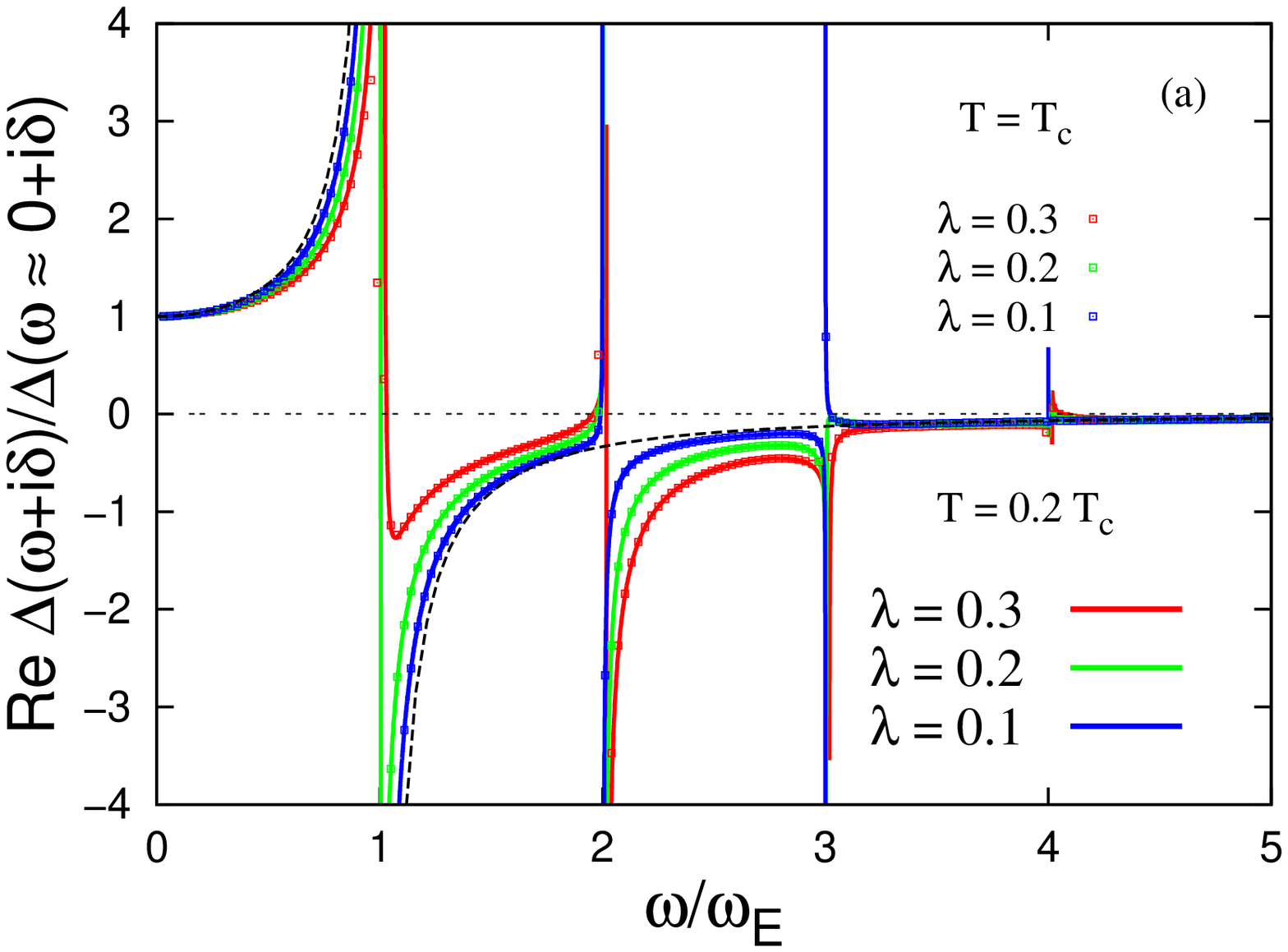}
\includegraphics[height=2.8in,width=2.4in,angle=-90]{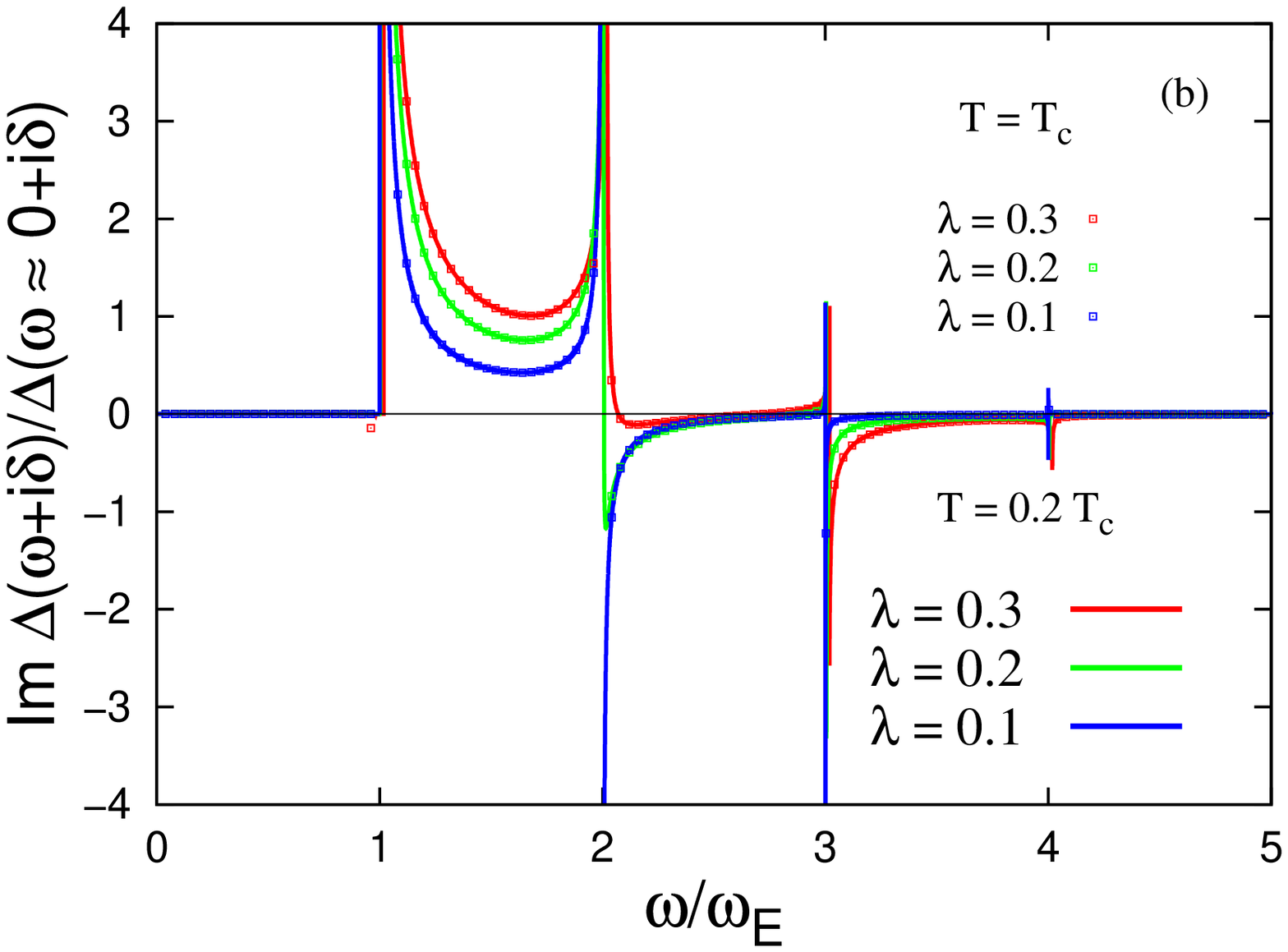}
\end{center}
\caption{(a) Real and (b) Imaginary parts of the {\it normalized}~\cite{remark3} gap function, $\Delta(\omega + i\delta)/\Delta(\omega \approx 0 + i\delta)$, at a low  temperature $T = 0.2T_c$, as a function of frequency, for the same three different coupling strengths, $\lambda = 0.3, 0.2, 0.1$.  These are shown as curves.  Also shown are the results at $T_c$ (naturally normalized) for the various coupling strengths, indicated with points of the same colour as their corresponding low temperature curves. 
They essentially match perfectly, indicating that the results are universal functions of frequency (independent of temperature) for a given coupling strength, as was the case in the imaginary frequency results.}
\label{fig6_mirabi}
\end{figure*}
To examine the frequency dependence of the gap function, we numerically solve Eqs.~(\ref{f1}-\ref{f2}) through an iterative process. Figure~(\ref{fig4_mirabi}) shows the
(a) real and (b) imaginary parts of the gap function as a function of frequency for three different coupling strengths, $\lambda = 0.3, 0.2, 0.1$. 
These plots were all obtained from the linearized gap equations, i.e., at $T=T_c$, and therefore the gap function has been normalized in each case to have unit value at the origin.~\cite{remark3} 
Both the real and imaginary parts of the gap function show a sequence of resonances at integer multiples of the Einstein phonon frequency. 
These resonances are very sharp because we have used an Einstein spectrum for the phonons; otherwise they would be considerably broadened. 
We have also shown the $\lambda = 0$ limiting result, which is $\Delta(\omega + i\delta) = 1/[1 - (\omega/\omega_E)^2]$, in the dashed black curve, and it is clear that the numerical results are trending towards this  result as $\lambda$ decreases.

Figure~(\ref{fig5_mirabi}) shows the (a) real and (b) imaginary parts of the gap function as a function of frequency for the same three coupling strengths, but now at a low temperature ($T/T_c = 0.2$) where the low frequency gap function amplitude is fully developed. 
The energy scale of the gap function varies over more than two orders of magnitude, as $\lambda$ varies between $0.1$ and $0.3$, and therefore this plot is of little use. 
The results for $\lambda = 0.1$ are not even visible on this scale. 
Therefore, in Fig.~(\ref{fig6_mirabi}) we show results for the same three coupling strengths, but with the gap function normalized to its value at zero frequency.~\cite{remark3} 
Also shown is the $\lambda=0$ result (dashed black curve). 
We have also included symbols to indicate the results at $T=T_c$ from Fig.~(\ref{fig4_mirabi}), which are almost indistinguishable from the low temperature curves. 
Of course, they are actually distinguishable, as the opening of a gap affects the excitation spectrum, including the gap function at higher frequencies.
However, for these values of $\lambda$ the movement of these excitation energies is of order $(\Delta_0/\omega_E)^2 \approx 10^{-4}$ for the largest value of $\lambda$ and much smaller still for the others.

\begin{figure*}[tp]
\begin{center}
\includegraphics[height=2.8in,width=2.4in,angle=-90]{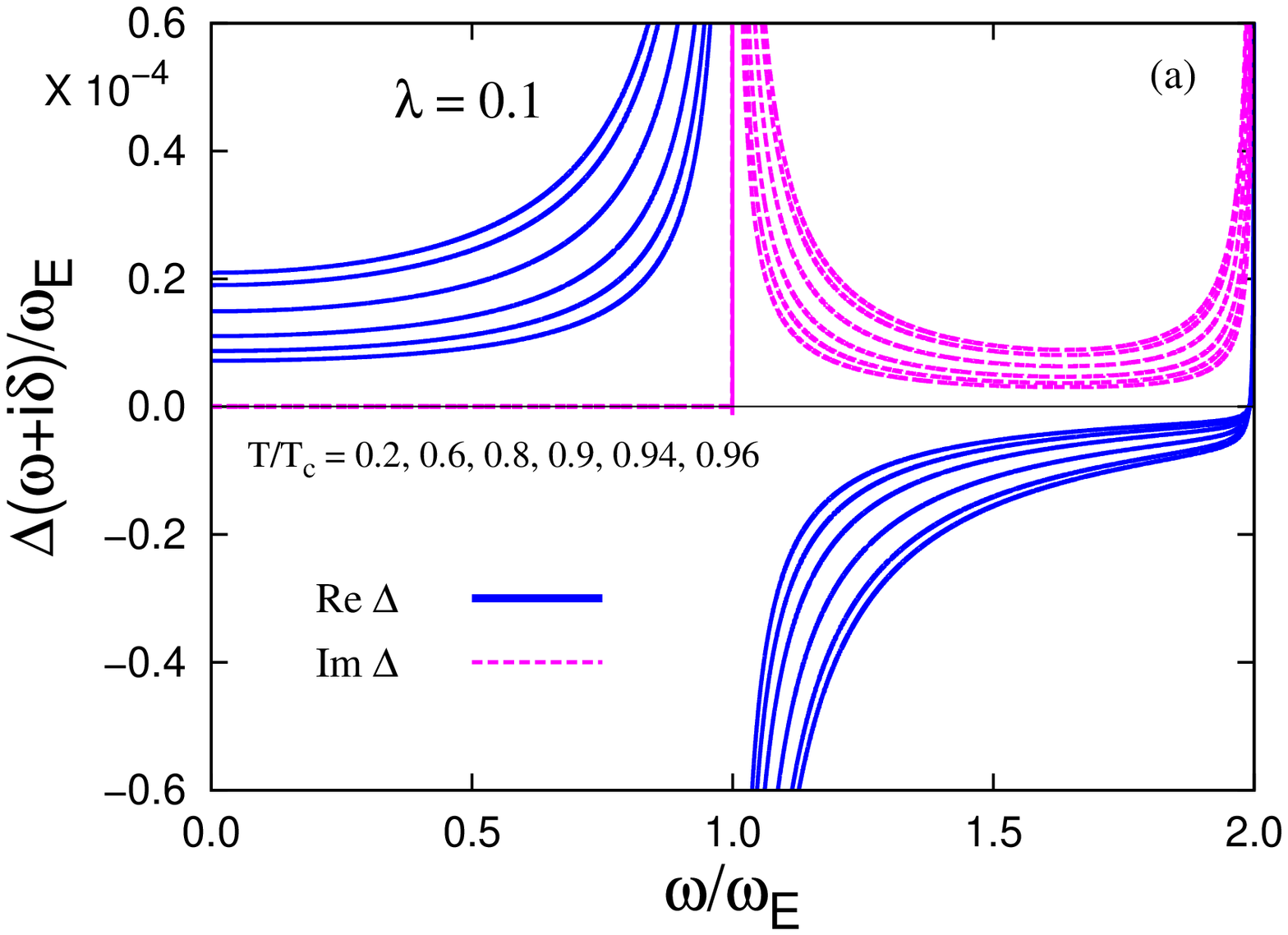}
\includegraphics[height=2.8in,width=2.4in,angle=-90]{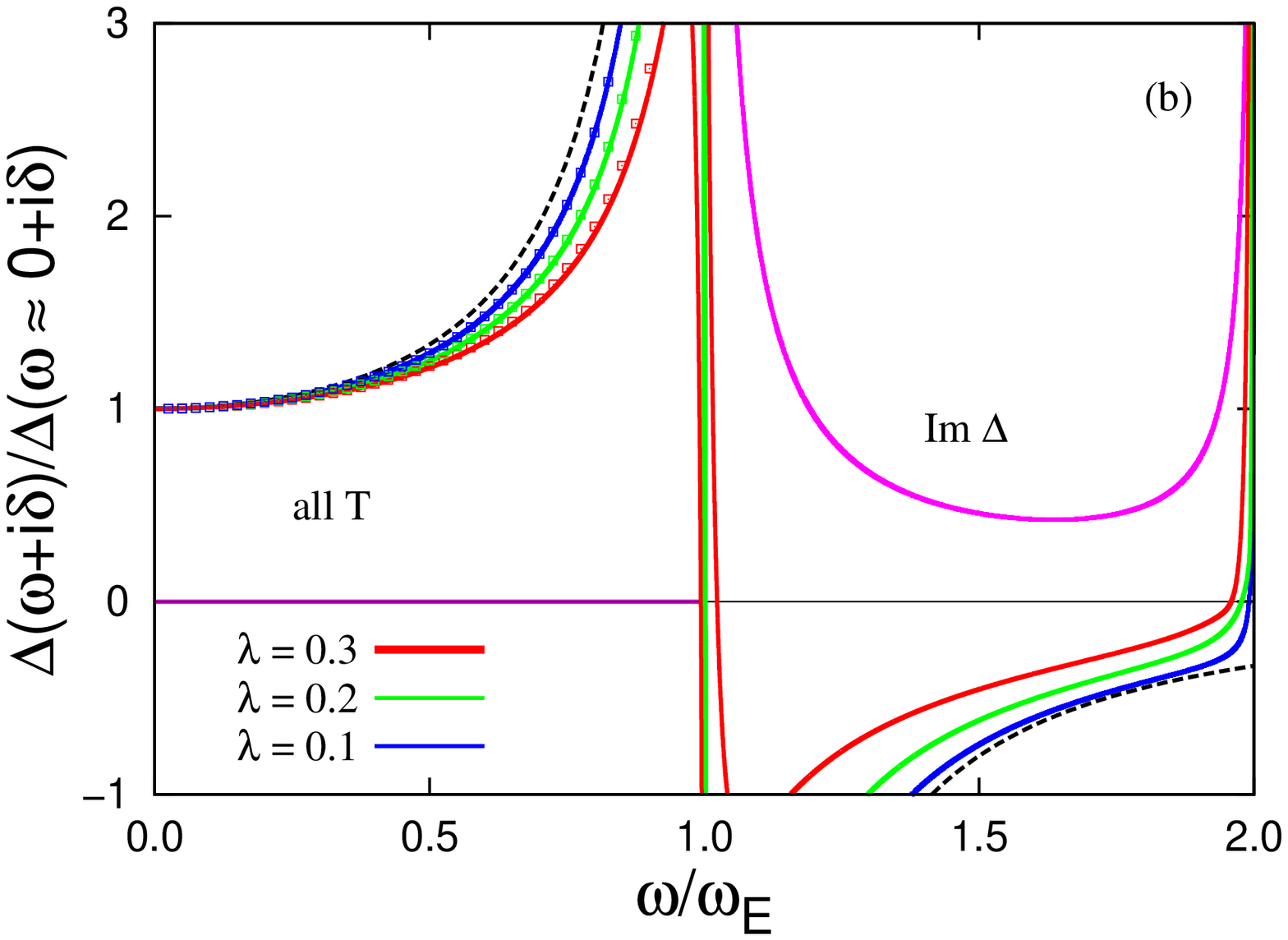}
\end{center}
\caption{(a) Real and (b) Imaginary parts of the gap function, $\Delta(\omega + i\delta)$, at various temperatures, as a function of frequency, specifically for $\lambda = 0.1$.
The temperatures used are as indicated, with higher amplitude curves corresponding to lower temperatures. 
In (b) we show the same data, now normalized to the low frequency gap function. Both real and imaginary parts collapse to a single curve, indicated in blue (real part) and purple (imaginary part). 
We also show the normalized results for the real part for $\lambda = 0.2$ (green curve) and $\lambda = 0.3$ (red curve), representative of any temperature below $T_c$ (we used the result at $T_c$ in each case for definiteness). 
We also plot the approximate expression, Eq.~(\ref{gapapp1}), for $\omega < \omega_E$, with points (using the same colour) for each value of $\lambda$. 
There is clearly excellent agreement, with slight deviations visible for $\lambda = 0.3$.}
\label{fig7_mirabi}
\end{figure*}

To further emphasize this scaling, we show the gap function for a variety of temperatures versus frequency in Fig.~(\ref{fig7_mirabi}) for $\lambda = 0.1$.
In (a) the gap function is plotted in units of the phonon frequency. 
For the real part (blue solid curves) the entire scale of the gap function gradually goes to zero as the temperature approaches the critical temperature. 
In (b) the gap function for each temperature is normalized to the zero-frequency value~\cite{remark3} of the real part for the same temperature [as we did in Fig.~(\ref{fig6_mirabi})]. 
It is quite striking how all the curves in part (a) collapse onto a single curve in (b), for both the real (solid blue curves) and imaginary (solid purple curves) parts. 
In (b) we have also included results for $\lambda = 0.2$ and $\lambda = 0.3$. 
We have also indicated in (b) the standard $\lambda \rightarrow 0$ result, $\Delta(\omega + i\delta)/\Delta(0) = 1/[1-(\omega + i\delta)^2]$, and it is clear that the numerical result for $\lambda = 0.1$ is quite close to this one. 
However, Eq.~(\ref{hfunc}) allows us to improve upon this. 
In fact, we only need to use the $T\ll\omega_{E}$ limit of Eq.~(\ref{hfunc}), given in Eq.~(\ref{hasym}), 
combined with the weak-coupling approximation for the renormalization, given in Eq.~(\ref{zzapp_real}), to obtain a simple approximate 
expression (valid to first order in $\lambda$) for the gap function via Eq.~(\ref{pert}):
\bwt
\begin{align}
\Delta^{(1)}_{\rm app}(\omega + i\delta) &={1 \over Z^{(1)}_{\rm app}(\omega + i\delta)} {\Delta_0 \over 1 - \bar{\omega}^2}\left\{1+\lambda\left[{3 \over 2}   - {1 \over 4 - \bar{\omega}^2}\left( {2 + \bar{\omega}^2 \over 2\bar{\omega}}{\rm \ell n}\left|  {1 + \bar{\omega}  \over 1 - \bar{\omega} } \right|  - {3 \over 2} {\rm \ell n}\left|1 - \bar{\omega}^2\right| \right)\right]\right\} , \quad \omega < \omega_E.
\label{gapapp1}
\end{align}
\ewt
This result is purely real, and is plotted with symbols in Fig.~(\ref{fig7_mirabi}) for the three values of $\lambda$,
and shows excellent agreement with the numerical results over the domain $0< \omega < \omega_E$.
\begin{figure}[tp]
\begin{center}
\includegraphics[height=2.8in,width=2.4in,angle=-90]{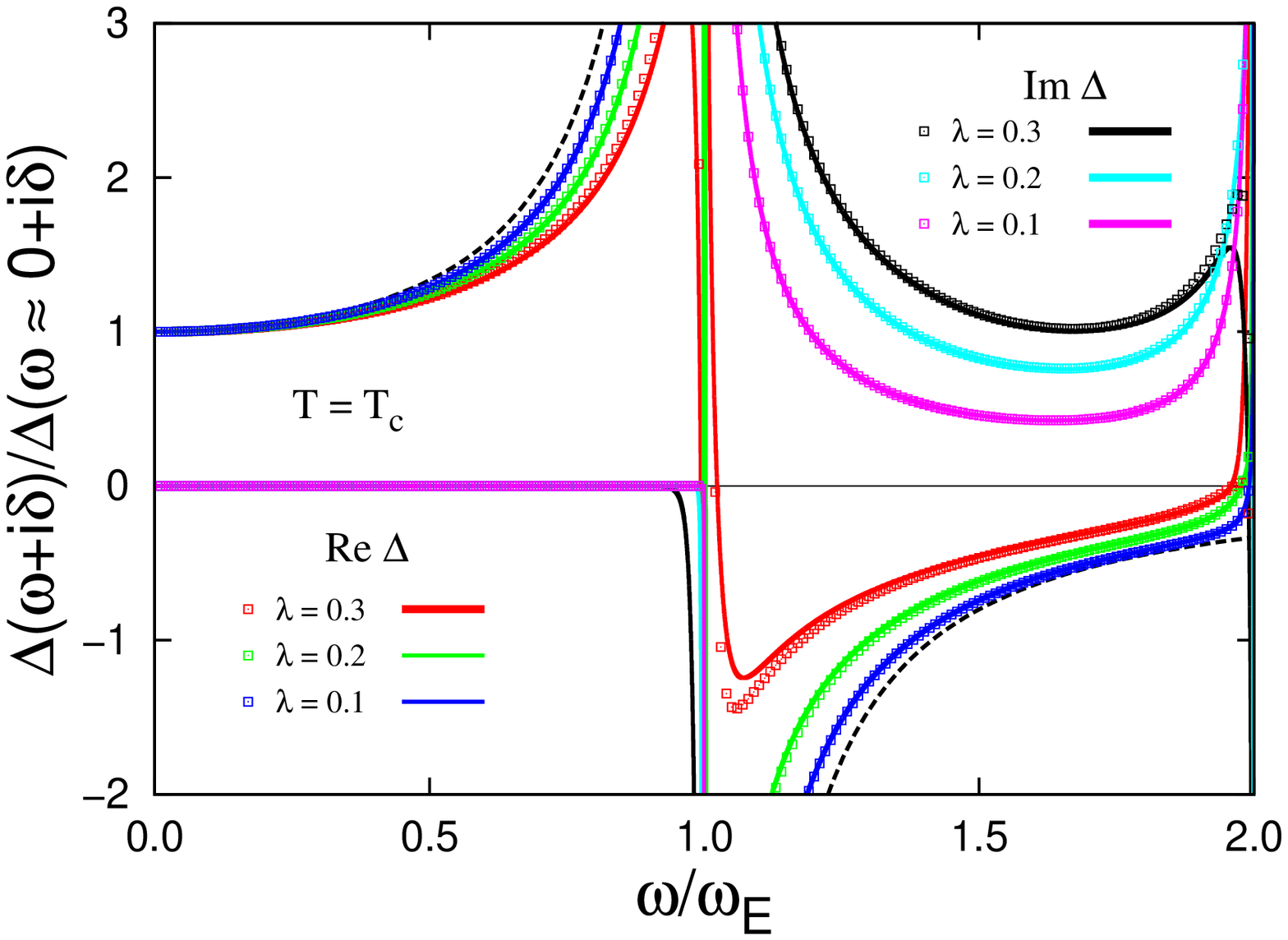}
\end{center}
\caption{Real and Imaginary parts of the normalized gap function, $\Delta(\omega + i\delta)/\Delta(\omega \approx 0 + i\delta)$, at $T_c$, as a function of frequency, for $\lambda = 0.1, 0.2, 0.3$. 
Points are now included corresponding to the approximation Eq.~(\ref{gapapp1}) for $\omega < \omega_E$, and to the approximation
Eq.~(\ref{gapapp2}) for $\omega_E < \omega < 2\omega_E$, with the same color as that used in the curves showing the full numerical results. 
The approximation in the extended frequency region is clearly excellent, for both the real and imaginary parts. 
Essentially, Eq.~(\ref{gapapp1}) provides a correction to the black dashed curve (for the real part) of order $\lambda$, and provides the entire result for the imaginary part. It is very accurate, particularly for the two lowest values of $\lambda$ shown.}
\label{fig8_mirabi}
\end{figure}

Finally, to understand these curves a little better, we focus on the gap function at $T=T_c$. Having established that the results at $T=T_c$ are identical
to those at low temperatures, because the temperature scale is so low compared to $\omega_E$ in the weak-coupling limit, we can proceed to analyze Eqs.~(\ref{f1}, \ref{f2}) in this limit. 
Then, focussing on positive frequencies only, for $\omega < \omega_E$, only the first line in each of Eqs.~(\ref{f1}, \ref{f2}) needs to be considered, and as we have
now established, Eq.~(\ref{gapapp1}) gives a very good approximation for this region. 
It therefore stands to reason that for $\omega_E < \omega < 2\omega_E$, the second lines in each of Eqs.~(\ref{f1}, \ref{f2}) will contribute, with a contribution to the gap equation of approximately
\bwt
\begin{equation}
i \pi \lambda{ \omega_E \over 2} {\Delta(\omega - \omega_E + i\delta) \over \sqrt{(\omega - \omega_E)^2 - \Delta^2(\omega - \omega_E + i\delta) }} \approx
i \pi \lambda{ \omega_E \over 2} {\Delta(\omega - \omega_E + i\delta) \over \omega - \omega_E}, \quad \omega > \omega_E,
\label{2nd_gap}
\end{equation}
and similarly to the renormalization equation,
\begin{equation}
i \pi \lambda{\omega_E \over 2} {\omega - \omega_E  \over \sqrt{(\omega - \omega_E)^2 - \Delta^2(\omega - \omega_E + i\delta) }} \approx
i \pi \lambda{\omega_E \over 2}, \quad  \omega > \omega_E.
\label{2nd_zz}
\end{equation}
Then, defining 
\be
\phi^{(1)}_{\rm app}(\omega + i\delta) \equiv Z^{(1)}_{\rm app}(\omega + i\delta) \Delta^{(1)}_{\rm app}(\omega + i\delta) = {\Delta_0 \over
1 - \bar{\omega}^2} \left(1 + \lambda h(\omega)\right),  
\label{phiapp}
\ee
as in Eq.~(\ref{hfunc}) with the use of Eq.~(\ref{zzapp_real}), we readily obtain the approximation applicable in this frequency regime:
\begin{equation}
\Delta^{(2)}_{\rm app}(\omega + i\delta) = {\phi^{(1)}_{\rm app}(\omega + i\delta)  + i {\pi \lambda \over 2} {\phi^{(1)}_{\rm app}(\omega -\omega_E + i\delta)  
\over Z^{(1)}_{\rm app}(\omega -\omega_E + i\delta)} {\omega_E \over \omega - \omega_E} \over
Z^{(1)}_{\rm app}(\omega + i\delta) + i \pi \lambda{\omega_E \over 2\omega}}, \quad 2\omega_{E}>\omega > \omega_E,
\label{gapapp2}
\end{equation}
\ewt
which now has real and imaginary parts. 
These results are plotted in Fig.~(\ref{fig8_mirabi}), along with the numerical results, and show excellent agreement for $\lambda = 0.1$ and even $0.2$. 
For $\lambda = 0.3$ the agreement is still remarkably good over most of the frequency range. 
In this latter case (recall that $T_c \approx 0.01\omega_E$), signatures of non-negligible temperature effects are clearly present, 
within a frequency range of a few percent of  $\omega_E$
consistent with the magnitude of the temperature. If we try to define a $\lambda = 0$ limiting result, we find
\be
\Delta^{(0)}_{\rm app}(\omega + i\delta) =  {\Delta_0 \over 1 - (\bar{\omega} + i\delta)^2},
\label{gap00}
\ee
which is purely real except for $\delta$-functions at $\omega = \pm \omega_E$. So the presence of any non-singular imaginary component of the gap function as
seen in these results occurs due to a finite (non-zero) value of $\lambda$. Similarly, the non-trivial structure obtained in the numerical solutions beyond 
$\omega = \omega_E$ occurs because of
the finite value of $\lambda$. In fact, it is clear that the perturbation theory in $\lambda$ requires an outward progression in frequency, a natural consequence of
the fact that higher frequency excitations require more scattering processes with the phonons, and hence higher powers of $\lambda$.

\section{Conclusions}
\label{Conc}

Although many solutions to the Eliashberg gap equations are readily available in the literature, even on the real frequency axis, the weak-coupling limit is an exception. 
In this paper we have remedied this deficiency, with a combination of analytical solutions (with $\lambda$ as an expansion parameter) and numerical solutions. 
We have found that, contrary to the experience of many researchers (and our own),
the numerical solution should be easier to obtain than it is, since the gap function obeys a universal frequency dependence for an electron-phonon
$\delta$-function spectrum with a given strength (universal with respect to temperature below the critical temperature). 
The gap function has a pronounced frequency dependence in the weak-coupling limit, diverging at the Einstein frequency $\omega_E$, 
and with decaying resonances thereafter, in both the real and imaginary parts. 
In fact, the expansion in $\lambda$ also organizes itself with increasing frequency -- every integer multiple of $\omega_E$ requires an extra power of $\lambda$.

This non-trivial frequency dependence results in an altered solution for the zero-temperature energy gap. We obtain the analytical result:
\begin{equation}
\Delta_0 = {2 \hbar \omega_E \over \sqrt{e}} \exp\left(-{1 + \lambda \over \lambda}\right),
\label{summ_eqn}
\end{equation}
which has precisely the same $1/\sqrt{e}$ factor first discovered in the $T_c$ equation.~\cite{karakozov76} Interestingly, the mathematics to obtain this result
is very different than that at $T_c$, but this result is inevitable since previous work had established that various dimensionless BCS ratios (like the gap ratio) are achieved in the weak-coupling limit of Eliashberg theory. 
In fact, we have confirmed that the free energy also contains a correction $(1/\sqrt{e})^2$, so that the specific heat jump and a few other dimensionless ratios also achieve the BCS result in the weak-coupling limit of Eliashberg theory. 
With these results, a more complete understanding of the weak-coupling limit of Eliashberg theory has been achieved.

\begin{acknowledgments}

This work was supported in part by the Natural Sciences and Engineering Research Council of Canada (NSERC). 
RB acknowledges support from the Department of Physics and the Theoretical Physics Institute at the University of Alberta. 

\end{acknowledgments}

\appendix
\numberwithin{equation}{section}
\section{Derivation of the zero-temperature gap}

\bwt
With $Z \approx 1 + \lambda$, inserting Eq.~(\ref{kernel}) into the gap equation Eq.~(\ref{f1}) then results in
\begin{equation}
(1 + \lambda) \Delta(\omega + i\delta) = {1 \over 1 - (\bar{\omega} + i\delta)^2} \lambda \pi \bar{T} \sum_{m=-\infty}^{+\infty} 
\left(1 - {\bar{\omega}_m^2 + 2 i \bar{\omega}_m \bar{\omega} \over 1 + (\bar{\omega}_m + i \bar{\omega})^2} \right){\Delta(i\omega_m) \over \sqrt{\bar{\omega}_m^2 + \bar{\Delta}^2(i\bar{\omega}_m)}}.
\label{app1}
\end{equation}
This implies the ansatz, Eq.~(\ref{pert}), which also applies on the imaginary axis, where, on the right-hand side we use the low (imaginary-axis) frequency form for 
$h(i\omega_m) \approx 1 + \lambda$, so that the zeroth-order equation is
\begin{eqnarray}
(1 + \lambda) && = \lambda \pi \bar{T} \sum_{m=-\infty}^{+\infty}  {1 \over 1 + \bar{\omega}_m^2}{1 \over \sqrt{ \bar{\omega}_m^2 + \left({\bar{\Delta}_0 \over 1 + \bar{\omega}_m^2}\right)^2}} 
- \lambda \pi T  \sum_{m=-\infty}^{+\infty}  {\bar{\omega}_m^2 + 2 i \bar{\omega}_m \bar{\omega} \over 1 + (\bar{\omega}_m + i \bar{\omega})^2} {1 \over  1 + \bar{\omega}_m^2} {1 \over \sqrt{ \bar{\omega}_m^2 + \left({\bar{\Delta}_0 \over 1 + \bar{\omega}_m^2}\right)^2}}.
\label{app2}
\end{eqnarray}
\ewt
In the first term on the right-hand side of this equation it is critical to keep $\Delta_0$ in the denominator; however, the extra $(1 + \bar{\omega}_m^2)$ in the denominator
of $\bar{\Delta}_0$ is not required (by the time it begins to change the term it is already no longer contributing compared to the frequency). 
On the other hand, in the second term, the entire ${\bar{\Delta}_0 \over 1 + \bar{\omega}_m^2}$ in the square-root is not required, as the singular piece as $\bar{\omega}_m \rightarrow 0$ is no longer
present -- it has been cancelled by a factor of $\bar{\omega}_m$ in the numerator. 
Moreover, in the second term, since $\bar{\Delta}_0 << 1$ in the weak-coupling limit, we can set $\bar{\omega} = 0$. Using the prescription~\cite{AGDBook}
\begin{equation}
T \sum_{m=-\infty}^{+\infty} \rightarrow \int_{-\infty}^{+\infty} {d\omega \over 2\pi}, \quad {\rm as} \ T \rightarrow 0,
\label{pres}
\end{equation}
then the first term on the right-hand side is
\begin{equation}
\lambda \int_0^\infty d \omega^\prime{1 \over 1 + {\omega^\prime}^2}{1 \over \sqrt{{\omega^\prime}^2 + \bar{\Delta}_0^2}}.
\label{app4}
\end{equation}
Integrating by parts (with $dv = dw^\prime/\sqrt{{\omega^\prime}^2 + \bar{\Delta}_0^2}$), then substituting $y = {\omega^\prime}^2$, and recognizing that
\begin{equation}
\int_0^\infty dy {{\rm \ell n} y \over (1 + y)^2} = 0,
\label{integral}
\end{equation}
we obtain ${\rm \ell n}(2/\bar{\Delta}_0)$ for this first term, in the limit $\bar{\Delta}_0\ll1$.
The second term requires the same substitution and contributes $-\lambda/2$. Thus, Eq.~(\ref{app2}) then becomes
\begin{equation}
{1+\lambda \over \lambda} = {\rm \ell n}\left({2 \omega_E \over \Delta_0}\right) - {1\over 2},
\label{app5}
\end{equation}
The solution for $\Delta_{0}$ is thus
\begin{equation}
\Delta_0 = {2 \omega_E \over \sqrt{e}} {\rm exp}\left(-{1+\lambda \over \lambda}\right).
\label{app5}
\end{equation}
The additional $1/\sqrt{e}$ arises because of the frequency dependence of the pairing interaction, as occurred at $T_c$.~\cite{karakozov76}


\begin{thebibliography}{99}
\bibitem{remark1} Although we made no great effort to optimize convergence, in the interest of minimizing the number of Matsubara frequencies required, we implemented the following procedure. 
We performed summations up to some maximum cutoff, $N_{\rm max}$, with $\omega_{\rm max} \equiv (2N_{\rm max} - 1)\pi T$, and then
assumed that the order parameter beyond this frequency was given by the weak-coupling expression, 
$\Delta(i\omega_m) = \Delta(i\omega_{\rm max})(\omega_E^2+{\omega}_{\rm max}^2)/(\omega_E^2 + {\omega}_m^2)$, which was used to analytically sum the remaining frequencies (up to $m\rightarrow \infty$).
If we had used a non-zero value for the direct Coulomb repulsion, this accommodation of the high-frequency regime would be slightly more complicated.
\bibitem{remark2} We use the phrase ``on the real axis'' to mean a frequency $\omega + i\delta$ where $\delta$ is an infinitesimal, so strictly speaking this is meant to
mean ``just above the real axis.''
\bibitem{remark3} Again, the work of Karakozov et al.~\cite{karakozov76} shows that strictly speaking this is impossible, since the gap function goes smoothly to zero at the
origin, at finite temperature. However, with these values of $\lambda$,
as long as we stay above frequency scales  $\approx 10^{-20} \omega_E$, the gap function will appear to be approaching a constant at the origin (which is clearly
the case in the figures), and it is this constant which is normalized to unity.
\end{thebibliography}
\end{document}